\documentclass[lettersize,journal]{IEEEtran}
\usepackage{amsmath,amsfonts}
\usepackage{algorithmic}
\usepackage{algorithm}
\usepackage{array}
\usepackage[caption=false,font=normalsize,labelfont=sf,textfont=sf]{subfig}
\usepackage{textcomp}
\usepackage{stfloats}
\usepackage{url}
\usepackage{verbatim}
\usepackage{graphicx}
\usepackage{cite}
\usepackage{orcidlink}
\usepackage{bm}

\begin{document}

\title{Physics-Informed Deep Unrolled Network for Portable MR Image Reconstruction}

\author{Efe Il\i cak \orcidlink{0000-0002-7062-018X}, 
Chinmay Rao \orcidlink{0000-0002-2472-2409}, 
Chlo\'e Najac \orcidlink{0000-0002-7804-2281}, 
Beatrice Lena \orcidlink{0000-0001-9283-5918}, 
Baris Imre \orcidlink{0009-0004-7501-7584}, 
Fernando Galve \orcidlink{0000-0002-1600-7341}, \\
Joseba Alonso \orcidlink{0000-0002-2721-1380}, 
Andrew Webb \orcidlink{0000-0003-4045-9732}, 
Marius Staring  \orcidlink{0000-0003-2885-5812} 
\thanks{Manuscript received September XX, 2025. This work was supported by the European Innovation Council grant 101136407. (corresponding author: Efe Il\i cak)}
\thanks{This work involved human subjects or animals in its research. Approval of all ethical and experimental procedures and protocols was granted by the Medical Ethics Committee Leiden The Hague Delft, and performed in line with the Declaration of Helsinki.}
\thanks{Efe Il\i cak, Chinmay Rao, Baris Imre, and Marius Staring are with the Division of Image Processing, Department of Radiology, Leiden University Medical Center, 2333ZA Leiden, Netherlands. (e-mail: e.ilicak@lumc.nl; c.s.rao@lumc.nl; b.imre@lumc.nl; m.staring@lumc.nl)}
\thanks{Chlo\'e Najac, Beatrice Lena, and Andrew Webb are with the C.J. Gorter Center for High Field MRI, Department of Radiology, Leiden University Medical Center, 2333ZA Leiden, Netherlands. (e-mail: c.f.najac@lumc.nl; b.lena@lumc.nl; a.webb@lumc.nl)}
\thanks{Fernando Galve, and Joseba Alonso are with the Instituto de Instrumentación para Imagen Molecular, CSIC, Universitat Politècnica de València, 46003 Valencia, Spain. (e-mail: fernando.galve@i3m.upv.es; joseba.alonso@i3m.upv.es)}
}
\markboth{2025}
{Shell \MakeLowercase{\textit{et al.}}: A Sample Article Using IEEEtran.cls for IEEE Journals}

\IEEEpubid{0000--0000/00\$00.00~\copyright~2025 }

\maketitle

\begin{abstract}
Magnetic resonance imaging (MRI) is the gold standard imaging modality for numerous diagnostic tasks, yet its usefulness is tempered due to its high cost and infrastructural requirements. Low-cost very-low-field portable scanners offer new opportunities, while enabling imaging outside conventional MRI suites. However, achieving diagnostic-quality images in clinically acceptable scan times remains challenging with these systems. Therefore methods for improving the image quality while reducing the scan duration are highly desirable. Here, we investigate a physics-informed 3D deep unrolled network for the reconstruction of portable MR acquisitions. Our approach includes a novel network architecture that utilizes momentum-based acceleration and leverages complex conjugate symmetry of \textit{k}-space for improved reconstruction performance. Comprehensive evaluations on emulated datasets as well as 47mT portable MRI acquisitions demonstrate the improved reconstruction quality of the proposed method compared to existing methods.  
\end{abstract}

\begin{IEEEkeywords}
MRI, portable MRI, accelerated MRI, image reconstruction, deep learning, cross-domain generalization.
\end{IEEEkeywords}

\section{Introduction}
\label{sec:introduction}
\IEEEPARstart{L}{ow}-field (LF) magnetic resonance imaging (MRI) ($B_0 < 0.1$T) has seen a recent resurgence of interest owing to its cost effectiveness, and the desire to expand MRI systems beyond hospitals and specialized centers \cite{Campbell-Washburn2023}. In particular, LF MRI\footnote{In this work, we adopt the following nomenclature per \cite{Campbell-Washburn2023}: ultra-low-field [0mT, 10mT], very-low-field (10mT, 100mT], and mid-field (100mT, 1T].} scanners based on permanent magnet designs allow portable systems which have emerged as a compelling alternative for accessible and cost-effective point of care (POC) imaging \cite{Ayde2024,Rowand2025}. Recent works have demonstrated the viability of imaging with mobile systems outside clinical settings, ranging from intensive care units to sporting events, or even at patients' homes 
\cite{Guallart-Naval2022,Deoni2022,Algarn2023,Schote2025}. 

However, achieving diagnostic-quality images at low-field strengths poses a significant challenge \cite{Webb2023}, thus limiting the adoption of VLF MRI. Fundamentally, the MRI signal is proportional to the square of the magnetic field strength \cite{Marques2019}, which results in a limited signal-to-noise ratio (SNR) due to low spin polarization. Furthermore, the reduced size of portable systems limits the achievable $B_0$ homogeneity and gradient performance \cite{Webb2023}. In addition, the dominant noise source at LF systems is the coil noise, as opposed to the body noise \cite{Wu2016}, which limits the benefits of phased array coils \cite{Ayde2024}. Therefore, most portable systems utilize a single solenoid receiver coil to improve sensitivity \cite{Webb2023}, while simultaneously improving the cost effectiveness and system portability. Consequently, portable VLF scanners suffer from low SNR levels with limited redundancy in the acquired data coupled with long scan times.

Over the past several decades, constrained reconstruction methods have been developed and integrated into clinical practice to improve MR image quality and to reduce scan time \cite{Haldar2020, Fessler2020}. These methods include partial Fourier, which leverages conjugate symmetry \cite{Haacke1991,Noll1991}; parallel imaging, which utilizes coil-specific local signal modulations in multi-channel receiver arrays \cite{Pruessmann1999, Griswold2002}; compressed sensing, which exploits compressibility of the data \cite{Lustig2007}; low-rank based methods which exploit the inherent redundancies in the data \cite{Shin2014, Haldar2014}; and approaches that combine these priors \cite{Lingala2011, Haldar2016, Ilicak2023}. While these methods have been successfully integrated at mid-field (MF) and high-field (HF) strengths, they are not readily applicable to imaging with portable LF systems, due to the inherently low SNR and hardware constraints. 

In addition to the classical iterative reconstruction methods, deep-learning (DL) based methods have been proposed to learn relevant features from the data, with the promise of outperforming classical techniques \cite{Lin2022,Hossain2024}. Among these, several recent works have focused on developing physics-informed models \cite{Zhang2018, Aggarwal2019, Pezzotti2020, Sun2024, Giannakopoulos2024}, that incorporate domain knowledge to establish a connection between the iterative methods and DL-based approaches, while achieving superior reconstruction performance \cite{Jiang2024}. Essentially, these methods unroll the steps of an iterative optimization algorithm as a network architecture \cite{Monga2021}. 
\IEEEpubidadjcol

While DL-based reconstructions have been extensively studied for HF MRI, only a few studies have investigated their application at LF MRI, particularly as unrolled networks \cite{Shimron2024}. A key challenge for DL-based methods at LF is the lack of large datasets necessary for effective model training \cite{Hou2025}. Although several large-scale MRI datasets are publicly available \cite{VanEssen2013,Zbontar2018,Lyu2023}, they consist of MR images obtained with MF and HF scanners using multi-coil receiver arrays. As a result, even if tissue contrast differences between field strengths are ignored, they are not readily suitable for training LF networks due to differences in SNR, coil sensitivity profiles, and phase offsets between the receiver coils \cite{Tygert2018}.

To address these drawbacks, here we propose a data processing framework together with a novel deep unrolled network model for image reconstruction. Our approach employs a linear coil combination method to emulate the response of a single-channel receiver while matching the in-plane resolution and SNR levels of a target portable MRI system. The emulated data are then used for the training of a 3D deep unrolled network that leverages the momentum term for accelerated convergence together with the conjugate symmetry property of \textit{k}-space for improving conditioning. The proposed model was evaluated on emulated \textit{in vivo} brain MR images and, importantly, assessed for its ability to generalize across a domain shift by reconstructing undersampled \textit{in vivo} acquisitions from a portable 47mT Halbach-based VLF system. Our contributions can be summarized as follows:
\begin{itemize}
    \item We develop a framework to generate realistic single-coil VLF MRI acquisitions from publicly available multi-coil HF datasets. Our approach employs a linear coil combination method to emulate the response of a single-channel coil, followed by \textit{k}-space cropping for downsampling, and complex Gaussian noise addition to match target VLF system SNR characteristics.
        
    \item We present a novel deep unrolled reconstruction network based on the accelerated gradient descent algorithm. Our model incorporates 3D ResNet blocks with pre-activation architecture, while enforcing the conjugate symmetry property of the \textit{k}-space, and integrates data consistency projections to maintain fidelity with measurements. 
    
    \item We evaluate the performance of the proposed network compared to state-of-the-art methods trained on publicly available data via the proposed data emulation framework, and critically, demonstrate its cross-domain generalization by applying it directly to undersampled VLF measurements, including acquisitions with unseen image contrasts.
\end{itemize}

\section{Methods}
\label{sec:methods}
\subsection{Problem Formulation}

Let $\bm{y} \in \mathbb{C}^{M}$ be the undersampled noisy \textit{k}-space data acquired by the MRI scanner, and $F_{\Omega} \in \mathbb{C}^{N}$ the encoding operator that includes the Fourier operator with the sampling operator $\Omega$. The forward model can be described as: 
\begin{equation}
\bm{y} = F_{\Omega}\bm{x} + \eta,
\label{Forward Model}
\end{equation}
where $\bm{x}$ is the ground-truth signal and $\eta$ is the complex-valued measurement noise. The goal of MR image reconstruction is the recovery of $\bm{x}$ from the measurements $\bm{y}$ \cite{Fessler2020}. Due to the ill-posed nature of this problem, the recovery can be cast as the following optimization problem:
\begin{equation}
\underset{\bm{x}}{\text{arg min}} \|F_{\Omega}\bm{x}-\bm{y}\|_2^2  + \lambda R(\bm{x}),
\label{Regularized Problem}
\end{equation}
where the first term represents data consistency, and the second term $R(\bm{x})$ denotes a regularization term, such as low-rank or sparsity promoting priors, and $\lambda$ is a hyperparameter that balances data consistency and regularization. 
This problem can be solved iteratively via the gradient descent method: 
\begin{equation}
    \bm{x}^{i+1} =  \bm{x}^{i} - \alpha \nabla f(\bm{x}^{i}),
\label{GradDesc}
\end{equation}
where $i$ denotes the iteration, $\alpha$ is the step size, and $f$ is the objective function to be minimized. The convergence behavior of gradient descent method can be further improved by introducing a momentum term $\bm{v}^i$. In the case of Nesterov's accelerated first-order method \cite{Nesterov1983}, rather than evaluating the gradient at $\bm{x}^i$, it is evaluated at $\bm{x}^i + \mu^i \bm{v}^i$, with $\bm{v}^i = (\bm{x}^i-\bm{x}^{i-1})$ and $\mu^i$ is the momentum coefficient. Accordingly, the iterations steps can be expressed as \cite{Sutskever2013}: 
\begin{equation}
    \bm{x}^{i+1} = \bm{x}^i + \mu^i \bm{v}^i - \alpha \nabla f(\bm{x}^{i}+\mu^i \bm{v}^i).
\label{Nesterov}
\end{equation}
In the case of physics-based models, the steps of gradient descent algorithms are unrolled to yield a deep network, with the gradient step being replaced with a neural network $N$. In the case of momentum accelerated networks, recent works have displayed improvements to convergence rate and gradient flow \cite{Hosseini2020a, Hosseini2020b, Xin2025}. Inspired by these, we propose the following unrolled model where the iterations can be expressed as:
\begin{equation}
    \begin{split}
        \bm{h} &= \bm{x}^i+\mu^i \bm{v}^i\\ 
        \bm{x}^{i+1} &= \bm{h} - \lambda^i  F_{\Omega}^H (F_{\Omega}\bm{h} - \bm{y} ) - N^{i}(\bm{h})  \\
        \bm{v}^{i+1} &= \bm{x}^{i+1} - \bm{x}^{i}
    \end{split}
\label{Proposed}
\end{equation}
where the resulting model incorporates a data consistency term at every iteration, and $N^i$, $\mu^i$, and $\lambda^i$ are the trainable parameters unique for each iteration block. 

\subsection{Implementation Details}
\begin{figure}
    \centering
    \includegraphics[width=0.45\textwidth]{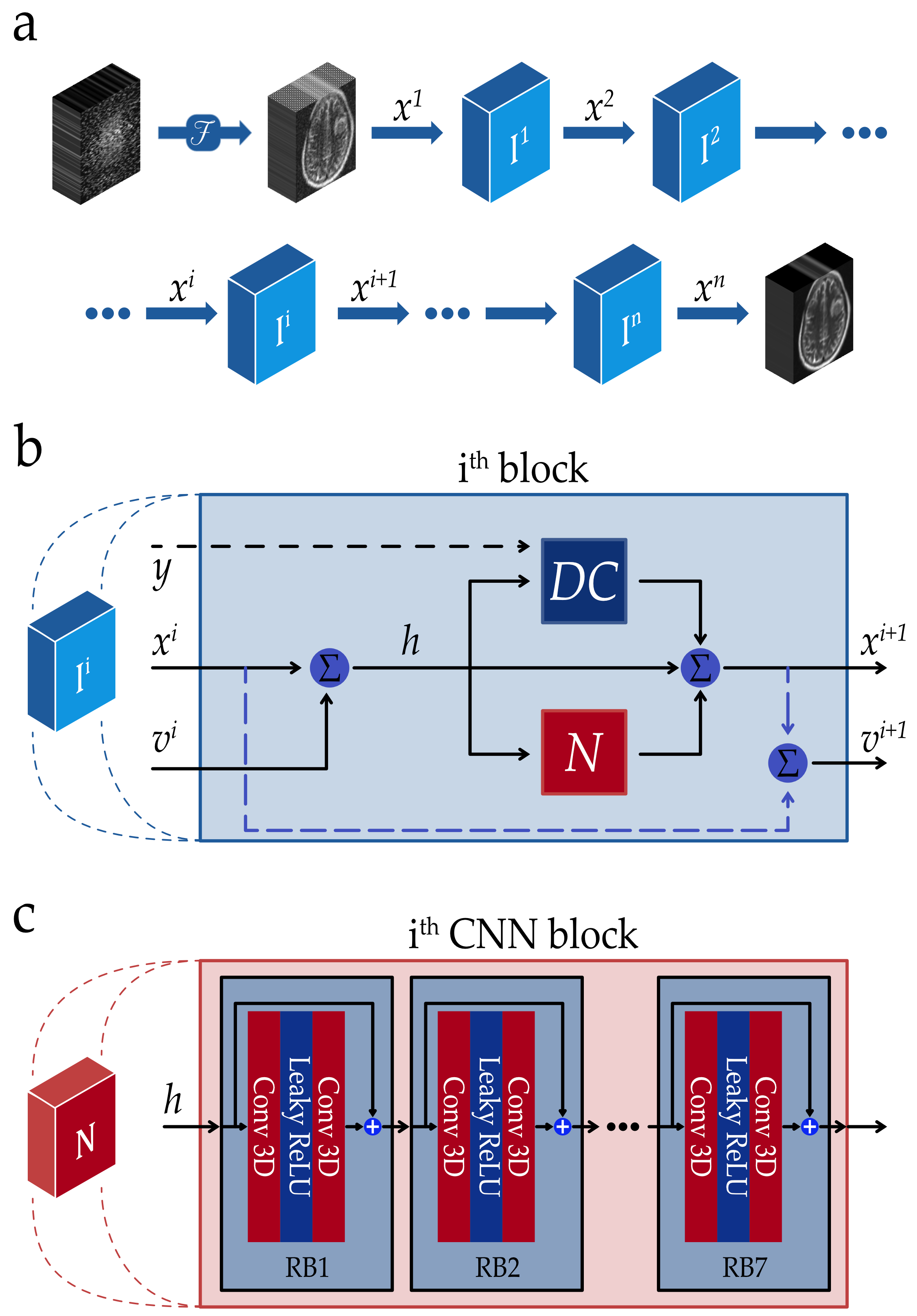}
    \caption{Overview of the network architecture. (a) The reconstruction is modeled as an unrolled network with fixed iterations. (b) In each iteration block, an accelerated gradient descent step is evaluated, which includes data consistency, residual connection, and a CNN projection. (c) Each CNN block is constructed using ResNet-based blocks (RBs) comprised of 3D convolutions interleaved with LeakyReLU activation functions.}
    \label{Fig:Arch}
\end{figure}

We implemented the unrolled network as a cascaded architecture consisting of five iteration blocks. Each iteration block includes a data consistency projection, a residual connection, and a convolutional neural network (CNN). A separate CNN was used for each block, each comprising seven ResNet-based convolutional blocks (RBs). All convolutions were three-dimensional with kernel sizes of $3\times3\times3$ and 32 channels, except for the input and output layers. Each RB follows the pre-activation design \cite{He2016} and uses LeakyReLU as the activation function. An overview of the deep unrolled network architecture is shown in Figure \ref{Fig:Arch}.

To improve the conditioning of the reconstruction problem, we employed the virtual conjugate coil (VCC) method \cite{Blaimer2009} to exploit the inherent complex-conjugate symmetry of \textit{k}-space. In the ideal case, the acquired MR signal $S(k)$ originating from the real-valued spin density satisfies the Hermitian symmetry property $S(\bm{k})=S^*(-\bm{k})$ \cite{Haacke1991}. In practice, background phase variations break this symmetry, but they can be estimated, e.g. from the \textit{k}-space center \cite{Kang2024}. The VCC method therefore augments the measured data by creating a virtual coil from the complex-conjugate of the actual coil, $S_{VCC}(\bm{k})=S^*(-\bm{k})$, with phase correction applied prior to combining the original and virtual coil images. The resulting augmented complex-valued tensor was then split into real and imaginary components and concatenated along the channel dimension before being passed into the network \cite{Kang2024}, resulting in a 4-channel input-output structure for the network. 

During training, we employed a composite loss function combining pixel-wise Huber loss and perceptual feature loss. The perceptual loss was calculated based on features extracted with a pretrained VGG16 model, using a weighted average of activations from layers prior to the first three max-pooling operations, as described in \cite{Ghodrati2019}. A single network was trained jointly on four different high-field MRI contrasts (FLAIR, T\textsubscript{1}-w, T\textsubscript{1}-w+Gd, T\textsubscript{2}-w). To account for the varying signal intensities across different MRI contrasts, the composite loss was normalized by the target volume energy, defined as the mean squared magnitude of the complex-valued ground truth data. Based on the ablation experiments, the final network had the following implementation details: 5 iteration blocks with 7 RBs at each iteration block, the VCC technique with phase correction, and the momentum acceleration implemented using learnable coefficients initialized from the FISTA formulation \cite{Beck2009}. The final model comprised of 1.56M parameters. All of the network trainings and reconstructions were performed on a workstation equipped with an NVIDIA RTX 4000 Ada GPU using the PyTorch framework. 

\subsection{Data Emulation Framework}
To generate a large-scale datasets for training a VLF reconstruction network, a number of works have proposed different approaches ranging from resampling magnitude-only HF MRI data to using computer simulated cross sections \cite{Zhou2022,Salehi2025}. Most related to our approach is the work of \cite{Shimron2024}, which proposed the use of a non-linear coil compression method together with \textit{k}-space cropping and complex additive white Gaussian noise. 

In this work, we propose a framework that utilizes publicly available high-field MRI datasets. We utilize the fastMRI brain dataset \cite{Zbontar2018}, which consists of raw multi-coil data acquired using 1.5T and 3T clinical systems. Here, the multi-coil data were coil combined using the emulated single coil (ESC) method, a linear combination method based on Hellinger distance \cite{Tygert2018}, mimicking the response of a single-coil receiver without the SNR benefits of multi-coil receivers. Unlike other coil combination methods such as the sum-of-squares or coil compression methods \cite{Uecker2014, Biyik2018}, this approach preserves phase information and avoids the spatially varying noise attenuation due to the compression effects \cite{Tygert2018}. Following coil combination, the complex-valued images were transformed into 3D \textit{k}-space using the Fourier transform and downsampled to the target in-plane resolution via \textit{k}-space cropping. Lastly, to simulate VLF acquisition conditions, additional complex Gaussian noise was added to match the SNR levels of VLF acquisitions \cite{Wu2016}.

The prepared datasets were partitioned into training, validation, and test sets containing 400, 100, and 80 subjects, respectively. During training, coil-combined and downsampled images served as the target ground truth, while the corresponding inputs were the ones further degraded with complex Gaussian noise and random undersampling. Retrospective undersampling was prescribed using two-dimensional variable-density masks, with 12\% of central \textit{k}-space fully sampled at an acceleration factor of $R=2$. A new sampling mask was generated independently for each subject at every training epoch.

\subsection{Portable MRI Experiments}
To show the performance of the proposed method, we acquired images from five healthy volunteers using a prototype portable 47mT MRI system based on Halbach-magnet design \cite{OReilly2021}. In vivo acquisitions were conducted in accordance with approval from the local ethics committee, which set a maximum scan duration of 60 minutes per subject. Written informed consent was obtained from all participants prior to imaging. All scans were acquired using 3D turbo spin echo (TSE) sequences with a spatial resolution of $1.5 \times 1.5 \times 5$ mm\textsuperscript{3}, elliptical \textit{k}-space coverage, and with matrix size of $150 \times 136 \times 38$. The subjects were scanned with inversion recovery (IR-T\textsubscript{1})-, proton density (PD)-, and T\textsubscript{2}-weighted protocols. The parameters for these scans were as follows:  
IR-T\textsubscript{1}-w images were acquired with a TE/TR of $20/1200$ ms, inversion time (TI) of $91$ ms, echo train length (ETL) of 8, and an acquisition time (TA) of  9:53 (min:sec); PD-w images were acquired with a TE/TR of $16/600$ ms, ETL of 8, and TA of 5:06; T\textsubscript{2}-w images were acquired with a TE/TR of $150/2500$ ms, ETL of 15, and TA of 10:52.

The portable  MRI acquisitions were not used in any part of network training and served solely for evaluation. For the evaluation of reconstruction methods, the fully-sampled acquisitions were retrospective undersampled at an acceleration factor of $R=2$ using two-dimensional variable-density masks, with 12\% of central \textit{k}-space fully sampled. In addition, the fully-sampled VLF MRI acquisitions (denoted as `Reference') were denoised using the BM4D algorithm to provide a higher-SNR reference (denoted as `Reference+'). Reconstruction performances of competing reconstruction methods were evaluated using both references, where the results based on the denoised references are indicated by the `+' sign. 

\subsection{Evaluation of Emulation Frameworks}
To quantitatively assess the fidelity of emulated datasets, we compared our ESC-based framework against a recently proposed data generation approach based on the ESPIRiT framework \cite{Shimron2024}. This evaluation was performed using T\textsubscript{2}-w images obtained using the Halbach-based VLF MRI system, alongside the T\textsubscript{2}-w high-field fastMRI datasets processed using both data emulation frameworks. Following the linear coil combination and ESPIRiT-based compression methods, the resulting single-coil datasets were downsampled to match the in-plane resolution of the portable MRI acquisitions. The datasets were then systematically evaluated using statistical intensity distribution analyses and perceptual similarity metrics. 

For statistical comparison, intensity histograms with 512 bins were computed over the normalized intensity range of each volume, capturing the global intensity distribution characteristics. The Jensen-Shannon (JS) distance was employed to quantify the similarity between the intensity distributions of the emulated and actual portable MRI acquisitions. This symmetric metric provides a measure of distributional differences, where lower values indicate higher similarity. For each of the five subjects, the JS distance was calculated against emulated datasets, and the results were summarized across the subjects.

To complement the intensity-based assessment, we evaluated perceptual similarity using feature representations extracted from a pretrained VGG16 model \cite{Simonyan2015}. To quantify structural and textural congruence at multiple levels of abstraction, feature maps were extracted from 15 central cross sections at three intermediate stages of the network, corresponding to layers immediately preceding the first three max-pooling operations. For each of the five subjects, cosine similarity scores were computed between the feature embeddings of emulated and reference cross sections at each stage, and summarized across the subjects to provide a hierarchical measure of perceptual fidelity.

\begin{figure}
    \centering
    \includegraphics[width=1\linewidth]{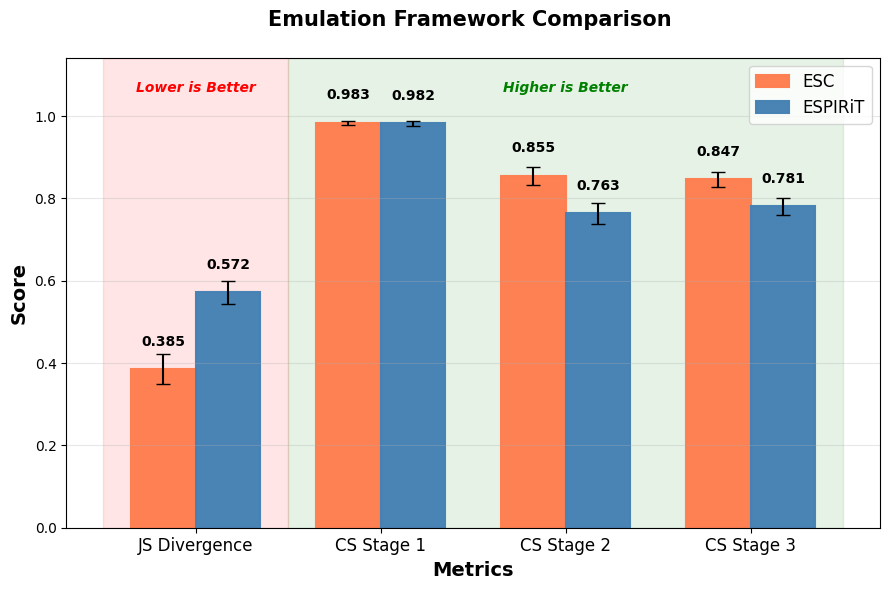}
    \caption{Performance comparison of data emulation frameworks evaluated using T\textsubscript{2}-weighted MRI images. Jensen-Shannon (JS) distance was computed to quantify intensity distribution differences, while cosine similarity (CS) was calculated to evaluate perceptual similarity at three distinct stages. The proposed framework based on the emulated single coil (ESC) method demonstrates superior performance with lower JS distance and consistently higher cosine similarity scores across all stages.}
    \label{Fig:EmulComp}
\end{figure}

\subsection{Ablation Experiments}
To assess the contributions of individual design components, we conducted a series of ablation studies. The first experiment evaluated the impact of different implementations of virtual conjugate coil technique. The second focused on the role of gradient momentum strategies. The third examined the effect of model capacity, by varying the number of unrolled iterations and residual blocks. 

\subsubsection{VCC Implementation}
To investigate the role of VCC in reconstruction performance, we trained and evaluated four variants of the proposed network: (i) without VCC, (ii) using VCC inputs with the network directly learning to fuse virtual and original coils, (iii) with conventional VCC without phase correction, and (iv) the conventional VCC with phase correction applied prior to combination (proposed). This comparison aimed to isolate the benefits of VCC formulation and its integration with the network. These experiments were conducted with a model comprised of 5 iterations, 7 RBs, and with learnable momentum terms starting from the same constant. 

\subsubsection{Momentum Implementation}
This experiment evaluated different strategies for incorporating momentum into the unrolled optimization process. Specifically, we compared: (i) no momentum, (ii) fixed coefficients based on the FISTA formulation \cite{Beck2009}, (iii) learnable momentum coefficients initialized from a constant, (iv) learnable momentum coefficients initialized from the FISTA formulation (proposed). The goal was to quantify the effect of adaptive momentum on reconstruction quality.

\subsubsection{Model Size}
To explore the effect of network depth and expressiveness, we varied the number of unrolled iterations and residual blocks (RB) per cascade. Configurations ranged from shallow (e.g., 3 iterations with 4 RBs) to deeper networks (e.g., 10 iterations with 7 RBs). This analysis aimed to characterize the trade-off between reconstruction performance and model complexity.

\begin{table}[t]
\setlength\tabcolsep{4pt} 
    \caption{Ablation experiments results of the proposed network. Results are grouped by virtual conjugate coil (VCC) implementation, momentum strategy, and model capacity.}
    \centering
    \begin{tabular}{|c|c|c|c|c|c|}
        \hline
        Configuration & Param. & Iter. & Depth & PSNR & SSIM  \\
        \hline \hline
        \multicolumn{6}{|c|}{\textit{Virtual Conjugate Coil (VCC)}} \\
        \hline
        No VCC        &  1.54M    & 5 & 7 & { }32.81$\pm$2.59* & 0.87$\pm$0.08  \\
        VCC via CNN   &  1.55M    & 5 & 7 & { }32.93$\pm$2.65* & 0.87$\pm$0.08  \\
        VCC w/o PC    &  1.56M    & 5 & 7 & { }33.08$\pm$2.74* & 0.87$\pm$0.09  \\
        \hline
        \multicolumn{6}{|c|}{\textit{Momentum Strategy}} \\
        \hline
        No Momentum    & 1.56M     & 5 & 7 & { }33.07$\pm$2.72* & 0.87$\pm$0.09  \\
        Learned Const. & 1.56M     & 5 & 7 & 33.10$\pm$2.73 & 0.87$\pm$0.09  \\
        Fixed FISTA    & 1.56M     & 5 & 7 & 33.11$\pm$2.74 & { }0.87$\pm$0.09* \\
        \hline
        \multicolumn{6}{|c|}{\textit{Model Capacity}} \\
        \hline
        XS   & 0.44M & 3  & 4  & { }32.36$\pm$2.48* & { }0.87$\pm$0.09* \\
        S    & 0.73M & 5  & 4  & { }32.87$\pm$2.72* & { }0.87$\pm$0.09*  \\
        L    & 2.39M & 5  & 10 & 33.10$\pm$2.73 & 0.87$\pm$0.09  \\
        XL   & 3.12M & 10 & 7  & { }33.07$\pm$2.70* & 0.87$\pm$0.09  \\
        \hline
        Proposed & 1.56M & 5 & 7 & 33.13$\pm$2.77 & 0.87$\pm$0.09 \\
        \hline
    \end{tabular}
    \label{tab:ablation_combined}
\end{table}

\subsection{Competing Methods}
We compared the proposed network against state-of-the-art classical (BM4D, CS-TV), generic DL-based (U-Net, ViT), and deep unrolled network-based (ISTA-Net, ISTA-Net-3D, DUN-U-Net) methods across the test datasets. All DL-based models were trained for 100 epochs using the Adam optimizer, with identical loss functions to ensure a fair comparison. The unrolled networks incorporated learnable data consistency weights to balance network-driven priors with measurement consistency. For methods where complex-valued operations are not supported, the complex-valued inputs were split into two channels as real and imaginary components. 

\textbf{BM4D}: A classical volumetric denoising method widely used in low-field studies \cite{Ayde2024} based on block matching with 4D filtering \cite{Maggioni2013}. To support complex-valued inputs, BM4D was applied separately to the real and imaginary components of the images. The noise standard deviation was adaptively estimated following the procedure in \cite{Khare2012} from the finest-scale Daubechies-4 wavelet coefficients. 

\textbf{CS-TV}: A compressed sensing reconstruction based on total variation regularization was implemented using a modified iterative clipping algorithm that supports complex-valued inputs \cite{Ilicak2017}. The regularization strength was automatically determined from local standard deviation maps as described in \cite{Ilicak2023}. Reconstructions were performed for 25 iterations, using soft data consistency weights linearly decreasing from 1.0 to 0.2.

\textbf{U-Net}: A 3D U-Net architecture \cite{Ronneberger2015} based on the MONAI implementation \cite{Cardoso2022}. The U-Net model consisted of 5-levels, downsampling achieved via max-pooling, upsampling achieved via convolution operations, with LeakyReLU activation functions. The model had 5.75M trainable parameters.

\textbf{ViT}: A vision transformer model \cite{Dosovitskiy2020} based on the MONAI implementation \cite{Cardoso2022}. The ViT model was implemented using a patch size of $4\times9\times9$, MLP dimension of 768, hidden layer dimension of 384, 12 layers, 8 heads, and a dropout rate of 0.1. To handle varying image sizes, the volumes were resampled to a fixed size of $40\times189\times189$ using trilinear interpolation \cite{Lin2022}. The model had 23.34M trainable parameters.

\textbf{ISTA-Net}: A 2D unrolled model based on the iterative shrinkage-thresholding algorithm \cite{Zhang2018} that has been recently proposed for LF MRI reconstruction \cite{Shimron2024}. The network consists of 10 unrolled iterations, each composed of a data consistency projection followed by 2D convolutional blocks interleaved with ReLU activation functions and soft-thresholding operation \cite{Shan2023}. The model had 0.20M trainable parameters.

\textbf{ISTA-Net-3D}: This variant extends ISTA-Net by replacing 2D convolutions with 3D convolutions while preserving the original unrolled structure. Each of the 10 iterations includes a 3D convolutional block interleaved with ReLU activations and soft-thresholding, mirroring the architecture of the 2D version. The model had 0.59M trainable parameters.

\textbf{DUN-U-Net}: A deep unrolled network was implemented using 3D U-Net architecture using residual blocks and soft-thresholding operations as inspired by \cite{Pezzotti2020}. The network has consisted of 4-levels, downsampling achieved via adaptive average pooling to fixed spatial resolutions, upsampling performed through trilinear interpolation, with LeakyReLU activation functions. Soft-thresholding layers are inserted after each encoder blocks and bottleneck to promote sparsity and suppress noise. The model had 6.98M trainable parameters.

Reconstruction performances were evaluated using PSNR and SSIM metrics. Statistical significances were assessed using the Wilcoxon signed-rank test, with significance indicated as (*) for $p<0.01$, reported separately for PSNR and SSIM scores. 

\section{Results}
\label{sec:results}
\subsection{Emulated Dataset Similarity}
The quantitative performance comparison between the emulation frameworks is presented in Figure \ref{Fig:EmulComp}. The Jensen-Shannon (JS) distance measurements demonstrate that the proposed ESC-based method achieves lower divergence from the actual portable VLF MRI data compared to the ESPIRiT-based approach, indicating superior intensity distribution fidelity with an improvement of 32.66\%. For perceptual similarity assessment, cosine similarity values at Stage 1 were comparable between methods. However, the proposed framework demonstrated consistently higher mean similarity scores across all three processing stages, with notable improvements of 11.99\% and 8.40\% at Stages 2 and 3, respectively. These findings indicate enhanced preservation of higher-level structural and textural features throughout the processing pipeline. Collectively, these quantitative metrics indicate that the proposed ESC-based emulation framework provides superior data fidelity, better capturing both intensity distributions and perceptual characteristics of actual portable MRI acquisitions compared to the ESPIRiT-based framework.


\subsection{Ablation Results}
Model performances across ablation experiments are listed in Table \ref{tab:ablation_combined}. Here, PSNR and SSIM scores are calculated for the whole test set as (mean±std) across subjects. Statistical significances were assessed against the proposed model (5 iteration blocks, 7 residual blocks per iteration, phase-corrected VCC formulation, and learnable momentum coefficients initialized from the FISTA formulation).

The results demonstrate the effectiveness of the proposed architecture. Enforcing conjugate symmetry improves reconstruction performance, with the conventional phase-corrected VCC approach yielding the largest PSNR gain. Among momentum strategies, learnable coefficients initialized via the FISTA formulation achieves the highest PSNR score. In terms of network capacity, increasing the number of iteration blocks and the depth of CNN blocks enhances performance up to a point, beyond which performance gains saturate. Overall, the proposed model consistently outperforms its ablated variants, supporting the contribution of each design component.

\subsection{Emulated Data Results}
\begin{figure*}
    \centering
    \includegraphics[width=\textwidth]{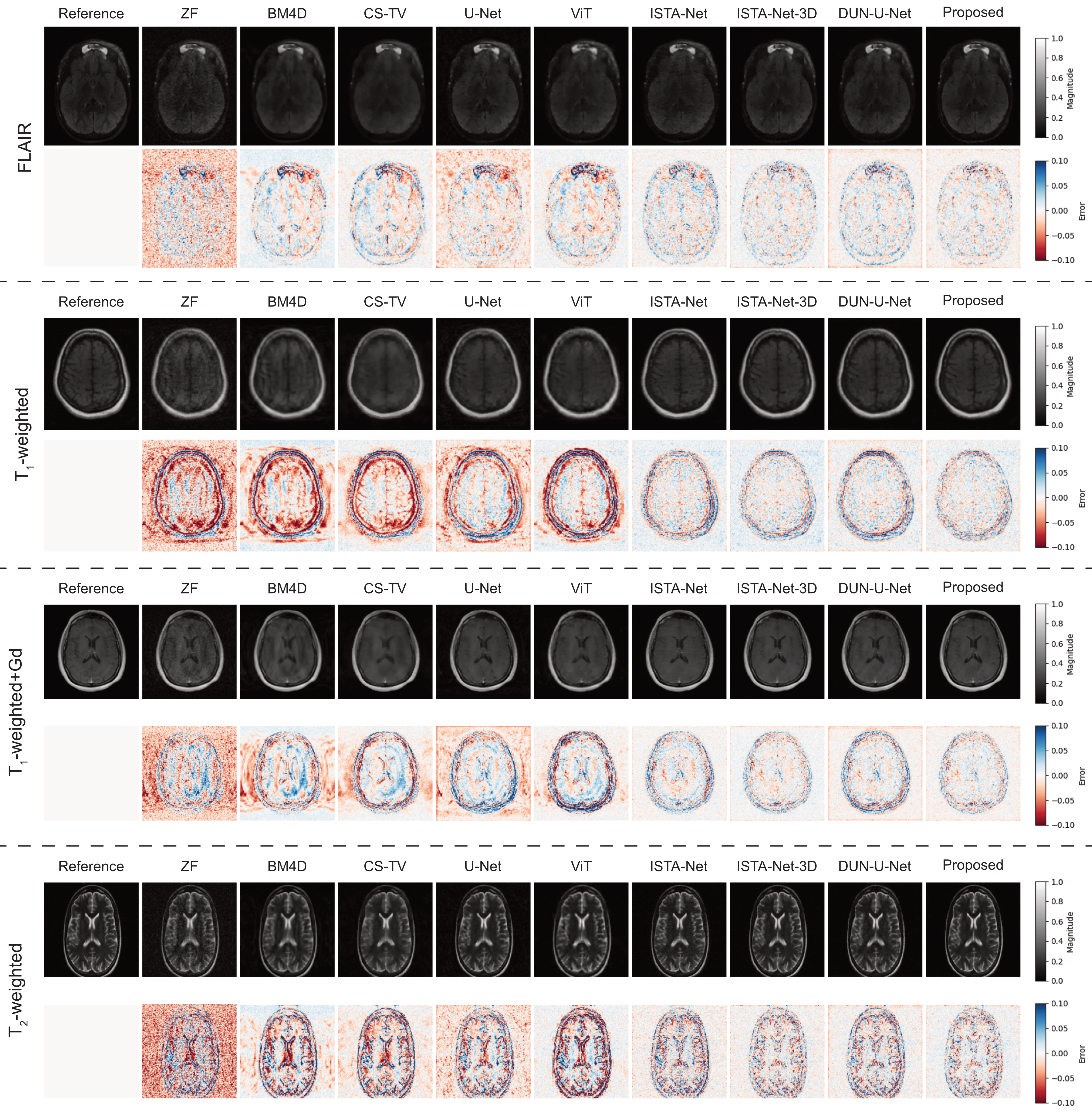}
    \caption{Representative reconstructions at $R=2$ for FLAIR, T\textsubscript{1}-w, T\textsubscript{1}-w+Gd, and T\textsubscript{2}-w from competing methods on the emulated VLF MRI test data, along with  references and zero-filled reconstructions (ZF). Error maps of each method are shown below with respect to the reference. Among the competing methods, the proposed method shows superior reconstruction quality with the least amount of residual reconstruction errors across the image contrasts.}
    \label{Fig:EmulTest}
\end{figure*}

For a fair comparison between different reconstruction methods, we evaluated all state-of-the-art approaches on the same emulated VLF MRI datasets. Representative reconstructions from all of the competing methods are displayed in Figure \ref{Fig:EmulTest} together with reference images and zero-filled (ZF) reconstructions. Among the classical methods, we observe that the BM4D denoised images suffer from aliasing artifacts resulting from undersampling. While the CS-TV is able to better suppress these artifacts, the reconstructions are overly smooth. Regarding the performance of generic DL-based reconstructions (U-Net and ViT), we observe that both methods perform worse than deep unrolled networks, despite having a larger number of learnable parameters. Among the deep unrolled networks (ISTA-Net, ISTA-Net-3D, DUN-U-Net, Proposed), we observe that the proposed method consistently outperforms the other reconstructions, achieving the highest image quality and artifact suppression. 

These observations are also reflected in the quantitative assessments listed in Table \ref{tab:Comparison}. Among the classical methods, we observe higher PSNR and SSIM scores with CS-TV method. Among the generic DL-based reconstructions, we observe that U-Net is able to achieve higher PSNR scores, whereas ViT is able to achieve higher SSIM scores. Nonetheless, both generic DL-based methods under-perform compared to deep unrolled networks despite their larger sizes. Among the deep unrolled networks, the proposed model achieves statistically significant improvements in both PSNR and SSIM compared to all other models. While our model has a larger parameter count than the second best performing model in terms of PSNR performance (ISTA-Net-3D with 0.59M trainable parameters), we also note that the S-variant of our architecture (with a comparable 0.73M trainable parameters), achieves higher PSNR and SSIM scores than ISTA-Net-3D on the same test set. Overall, the proposed method consistently outperforms the state-of-the-art methods.  
\begin{table}[]
    \caption{Reconstruction performance of different methods on the emulated data test sets. Best values for each metric are marked in bold, the second best values are underlined.}
    \centering
    \begin{tabular}{|c|c|c|c|c|c|}
    \hline
        Method & Param. & PSNR & SSIM & Infer. (s)  \\
        \hline
        \hline
        BM4D       & ---   &  29.08±2.08*     &    0.76±0.10*    &  5.08±0.18 \\
        CS-TV       &  ---   &  29.41±2.02*     &    0.81±0.08*   &  5.81±0.56  \\
        U-Net       & 5.75M     & 29.61±2.04*    &   0.74±0.07*    &  0.03±0.01 \\
        ViT        & 23.34M   &  29.18±1.83*     &   0.82±0.09*     & 0.01±0.00 \\
        ISTA-Net    &  0.20M  &  31.55±2.54*     &    0.83±0.10*    & 0.96±0.02 \\
        ISTA-Net-3D  &  0.59M  &  \underline{32.72±2.65}*     &    0.85±0.09*  &  0.08±0.02 \\
        DUN-U-Net     &  6.95M  &  31.74±2.29*     &    \underline{0.86±0.09}* &  0.19±0.01 \\
        Proposed   &  1.56M  &  \textbf{33.13±2.77}{ }     &   \textbf{0.87±0.09}{ }  & 0.35±0.04  \\ 
    \hline
    \end{tabular}
    \label{tab:Comparison}
\end{table}
        
\subsection{Portable MRI Results}
For the evaluation of reconstruction methods on the retrospectively undersampled portable VLF MRI acquisitions, we selected the top performing methods based on the emulated data experiments, representing the classical (CS-TV), generic DL-based (U-Net), and deep unrolled network (ISTA-Net-3D) approaches along with the proposed method. The reconstruction performances of competing methods were evaluated using both the original and the denoised references, where the results based on the denoised references are indicated by the `+' sign. 

\begin{figure*}
    \centering
    \includegraphics[width=\textwidth]{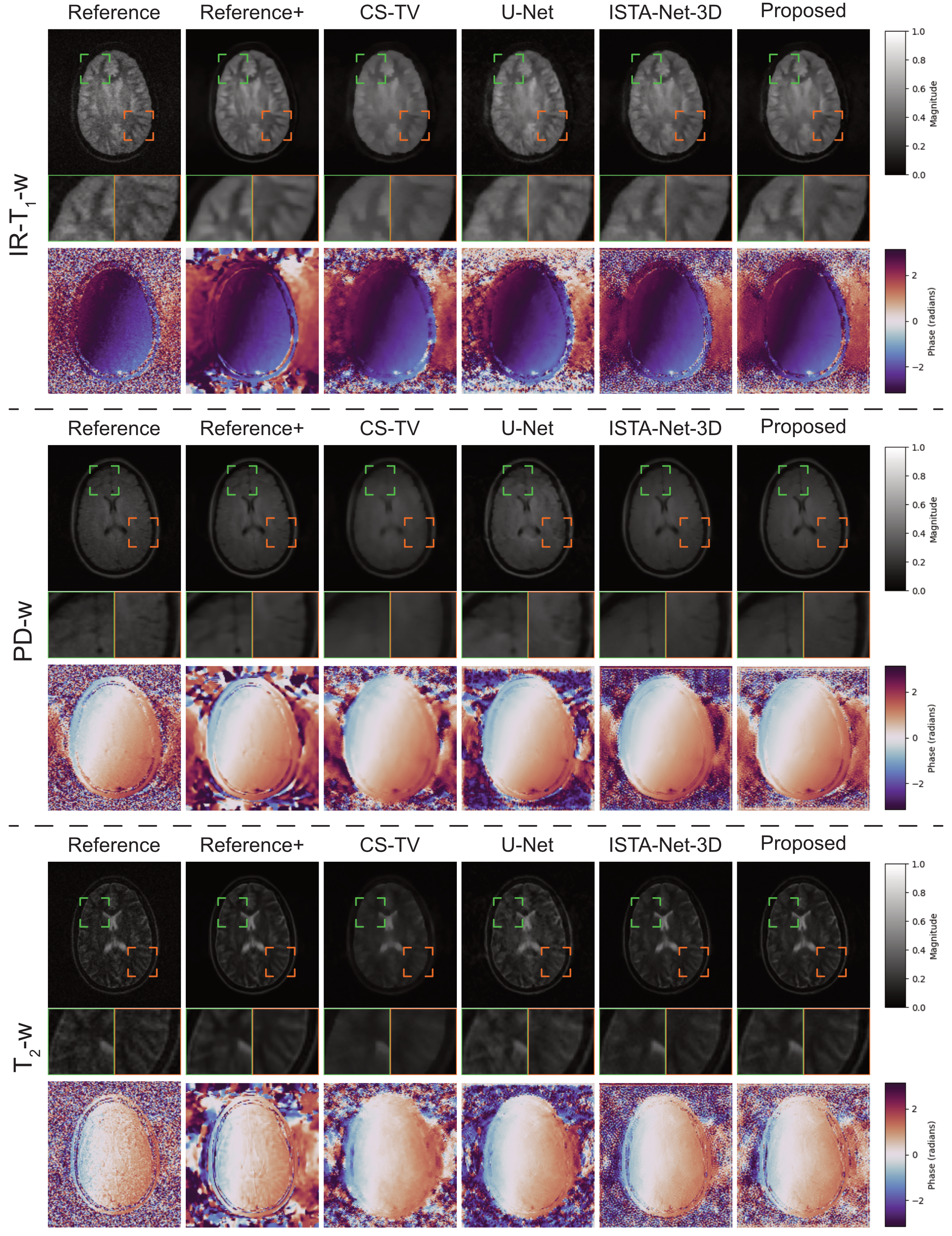}
    \caption{Representative magnitude and phase reconstructions of portable VLF MRI acquisitions at $R=2$ across different image contrasts, alongside fully-sampled references (`Reference') and denoised fully-sampled references (`Reference+'). The green and orange boxes depict regions that are shown in the zoomed-in panels below. The proposed methods retains fine tissue structures while effectively suppressing residual artifacts and noise, and produces smoother phase maps.}
    \label{Fig:Halbach}
\end{figure*}

Representative magnitude and phase reconstructions obtained from the 47mT Halbach-based VLF MRI system at $R=2$ are shown in Figure \ref{Fig:Halbach}, alongside the fully-sampled reference ('Reference') as well as BM4D-denoised fully-sampled reference ('Reference+'). Similar to the emulated dataset results, the CS-TV method produces overly smooth reconstructions, while the generic U-Net fails to adequately suppress aliasing artifacts. In contrast, both ISTA-Net-3D and the proposed method yield improved reconstructions. Notably, we observe that the proposed method is better able to recover finer structural details, while yielding smoother phase maps.

These visual observations regarding the performance on portable MRI datasets are also reflected in the quantitative results listed in Table \ref{tab:HalbachComparison}. With respect to both the original and denoised references, the proposed method consistently ranks among the top two performers, often achieving the highest PSNR and SSIM scores. Taken together, these results demonstrate the performance of the proposed deep unrolled network in reconstructing VLF MRI acquisitions, despite being trained only on emulated datasets with different image contrasts. 

\begin{table}[]
\setlength\tabcolsep{3pt} 
    \caption{Reconstruction performance of different methods across the portable VLF MRI acquisitions at $R=2$. Best values for each metric are marked in bold, the second best are underlined. PSNR+ and SSIM+ list the performances with respect to denoised references.}
    \centering
    \begin{tabular}{|c|c|c|c|c|c|}
    \hline
        Contrast & Method & PSNR & SSIM & PSNR+ & SSIM+  \\
        \hline
        \hline
                & CSTV                      &  26.47±1.85  &  0.52±0.03  & 33.23±1.27  &  0.82±0.01     \\
        IR-T\textsubscript{1}-w & U-Net     &  \underline{26.72±1.72}  &  \textbf{0.61±0.03}  & 27.48±1.61  &  0.58±0.04     \\
                & ISTA-Net-3D               &  26.37±1.79  &  0.48±0.05  & \underline{34.77±1.68}  &  \textbf{0.86±0.02}    \\
                & Proposed                  &  \textbf{26.77±2.22}  &  \underline{0.53±0.03}  & \textbf{35.08±1.32}  &  \underline{0.86±0.02}     \\ 
        \hline
                & CS-TV                     &   33.25±3.29  &  0.81±0.04  & 35.26±2.62  &  \underline{0.93±0.01}  \\
        PD-w    & U-Net                     &   31.51±3.00  &  \underline{0.84±0.05}  & 31.44±2.88  &  0.77±0.03  \\
                & ISTA-Net-3D               &   \underline{34.08±3.49}  &  0.73±0.02  & \underline{37.58±2.68}  &  0.92±0.01  \\
                & Proposed                  &   \textbf{35.35±3.86}  &  \textbf{0.85±0.04}  & \textbf{38.16±3.02}  &  \textbf{0.95±0.01}   \\ 
        \hline
                                & CS-TV     &   28.71±2.40  &  0.58±0.03  & 33.31±1.70  &  0.85±0.03           \\
        T\textsubscript{2}-w    & U-Net     &   28.89±2.17  &  \textbf{0.70±0.03}  & 29.95±1.89  &  0.66±0.04       \\
                                & ISTA-Net-3D&  \underline{28.94±2.44}  &  0.54±0.04  & \textbf{35.12±1.85}  &  \textbf{0.90±0.03}     \\
                                & Proposed   &  \textbf{29.61±2.84}  &  \underline{0.63±0.04}  & \underline{35.12±1.56}  &  \underline{0.87±0.01}    \\ 
        \hline

    \end{tabular}
    \label{tab:HalbachComparison}
\end{table}

\section{Discussion}
 
Prior works have proposed network-based reconstruction methods for low-field MRI acquisitions. Schlemper et al. \cite{Schlemper2019} proposed an unrolled network for non-Cartesian acquisitions, and Zhou et al. \cite{Zhou2022} further demonstrated this method for measurements obtained using a 64mT scanner. However, both used 2D U-Net-based architectures, whereas our model targets Cartesian data for its prevalence and robustness to system imperfections \cite{Ayde2024}. Furthermore, we propose a 3D network based on ResNet structure, using momentum terms, conjugate symmetry, and combined pixel-wise and perceptual loss. In \cite{Shimron2024}, the authors presented an unrolled network for the reconstruction of accelerated ultra-low-field MRI, and demonstrated their method using a 6.5mT scanner. Similar to the presented work, they used fastMRI dataset to generate single-channel training data. Yet, the authors employed ESPIRiT \cite{Uecker2014} method to obtain single-channel images, which inherently incorporates spatial denoising through the use of eigenvalue thresholding, resulting in unrealistic noise characteristics and phase behavior. Beyond differences in data preparation, our method leverages a more sophisticated network architecture by incorporating stronger prior information to improve reconstruction quality, and integrates data-consistency projections at every step to remain faithful to the acquired measurements. 

In addition to image reconstruction literature, some studies have recently proposed to use DL-based image translation networks to enhance the image quality. In \cite{Man2023}, the authors proposed a network with attention blocks to improve the spatial resolution and tissue contrast of low-resolution noisy ZF reconstructions, by learning priors from HF magnitude images; whereas in \cite{Lucas2023,Islam2023,Hsu2025}, authors have proposed methods using generative adversarial models for image-to-image translation. While the resulting images display improved spatial features and tissue contrast, the use of generative models and the omission of data consistency terms may result in substantial deviations from the actual measurements \cite{Cohen2018}, resulting in hallucinations and misleading features \cite{Rahman2021}. As opposed to these methods, our model utilizes the scanner measurements at every step to ensure data consistency. Nonetheless, similar image enhancement models can be merged with the method presented in this work to further enhance the image quality. 

The reconstruction performance of the proposed method can be improved further in several ways. Recent studies have shown that using pre-training strategies can improve the reconstruction performance of both CNN- and ViT-based networks, especially in limited data settings \cite{Mei2022, Lin2022}. Beyond population-based scan-general priors, recent studies have also investigated additional priors to improve the reconstruction quality. Subject-specific priors, such as acquisitions with different contrast weightings or prior scans of the same subject, have been explored to enhance reconstruction performance in \cite{Dar2020, Rao2024, Atalk2025, Oved2025}. In parallel, methods that leverage scan-specific priors, including dependencies in fully-sampled calibration regions \cite{Hosseini2020c}, have been proposed, together with hybrid approaches that combine scan-general and scan-specific information \cite{Dar2023}. While this study does not explore these additional priors, integrating them into the network could further improve reconstruction accuracy and robustness. 

\section{Conclusion}
In this work, we proposed a framework to emulate portable VLF MRI acquisitions from publicly available HF MRI acquisitions, and used these datasets to train a novel model for the reconstruction of portable VLF MRI acquisitions. In contrast to existing methods, our data emulation framework retains a better statistical and perceptual similarity to VLF measurements. Furthermore, we developed a physics-guided deep unrolled network that utilizes momentum-based acceleration while leveraging the complex conjugate symmetry property of the \textit{k}-space. Evaluations on both emulated datasets and acquisitions from a portable 47mT MRI scanner demonstrate the effectiveness of the proposed training framework. On the emulated datasets, the proposed model achieves superior reconstruction quality compared to state-of-the-art methods, while on portable MRI measurements, it attains comparable results when applied directly under a domain shift without retraining.



 





\bibliographystyle{IEEEtran}
\bibliography{IEEE_TCI}

\end{document}